\documentclass[letterpaper]{article} %
\usepackage{aaai2026}  %
\usepackage{times}  %
\usepackage{helvet}  %
\usepackage{courier}  %
\usepackage[hyphens]{url}  %
\usepackage{graphicx} %
\usepackage{natbib}  %
\usepackage{caption} %
\usepackage{algorithm}
\usepackage{algorithmic}

\usepackage{amsmath}
\usepackage{amssymb}
\usepackage{multirow}
\usepackage{enumitem}
\usepackage{bm}
\usepackage{xspace}
\usepackage{marvosym}
\usepackage{adjustbox}
\usepackage{booktabs}
\usepackage{array}
\usepackage{subcaption}
\usepackage[noabbrev,capitalise]{cleveref}

\usepackage{bbding}
\usepackage{tabularray}
\usepackage{makecell}
\usepackage{tikz}
\usepackage[table]{xcolor}
\usetikzlibrary{decorations.pathreplacing}

\usepackage{newfloat}
\usepackage{listings}
\DeclareCaptionStyle{ruled}{labelfont=normalfont,labelsep=colon,strut=off} %
\floatstyle{ruled}
\newfloat{listing}{tb}{lst}{}
\floatname{listing}{Listing}
\title{Inductive Generative Recommendation via Retrieval-based Speculation}
\author{
    Yijie Ding,\ \ Jiacheng Li,\ \ Julian McAuley,\ \ 
    Yupeng Hou\thanks{Corresponding author.} \\
}
\affiliations{
    University of California, San Diego\\
    \texttt{\{yid016, j9li, jmcauley, yphou\}@ucsd.edu}
}

\newcommand{\ie}{\emph{i.e.,}\xspace}
\newcommand{\eg}{\emph{e.g.,}\xspace}

\newcommand{\paratitle}[1]{\vspace{1.5ex}\noindent\textbf{#1}}
\newcommand{\wrt}{w.r.t.\xspace}
\newcommand{\ignore}[1]{}

\definecolor{dark2green}{rgb}{0.1, 0.65, 0.3}
\definecolor{dark2orange}{rgb}{0.9, 0.4, 0.}
\definecolor{dark2purple}{rgb}{0.4, 0.4, 0.8}

\begin{document}

\maketitle

\begin{abstract}
Generative recommendation (GR) is an emerging paradigm that tokenizes items into discrete tokens and learns to autoregressively generate the next tokens as predictions. 
While this token-generation paradigm is expected to surpass traditional transductive methods, potentially generating new items directly based on semantics, we empirically show that GR models predominantly generate items seen during training and struggle to recommend unseen items.
In this paper, we propose SpecGR, a plug-and-play framework that enables GR models to recommend new items in an inductive setting. SpecGR uses a \emph{drafter} model with inductive capability to propose candidate items, which may include both existing items and new items. The GR model then acts as a \emph{verifier}, accepting or rejecting candidates while retaining its strong ranking capabilities.
We further introduce the guided re-drafting technique to make the proposed candidates more aligned with the outputs of generative recommendation models, improving verification efficiency. We consider two variants for drafting: (1) using an auxiliary drafter model for better flexibility, or (2) leveraging the GR model's own encoder for parameter-efficient self-drafting. Extensive experiments on three real-world datasets demonstrate that SpecGR exhibits both strong inductive recommendation ability and the best overall performance among the compared methods. 
\end{abstract}

\begin{links}
    \link{Code}{https://github.com/Jamesding000/SpecGR}
\end{links}

\section{Introduction}\label{sec:intro}

\begin{figure}[t]
  \centering
  \includegraphics[width=0.95\linewidth]{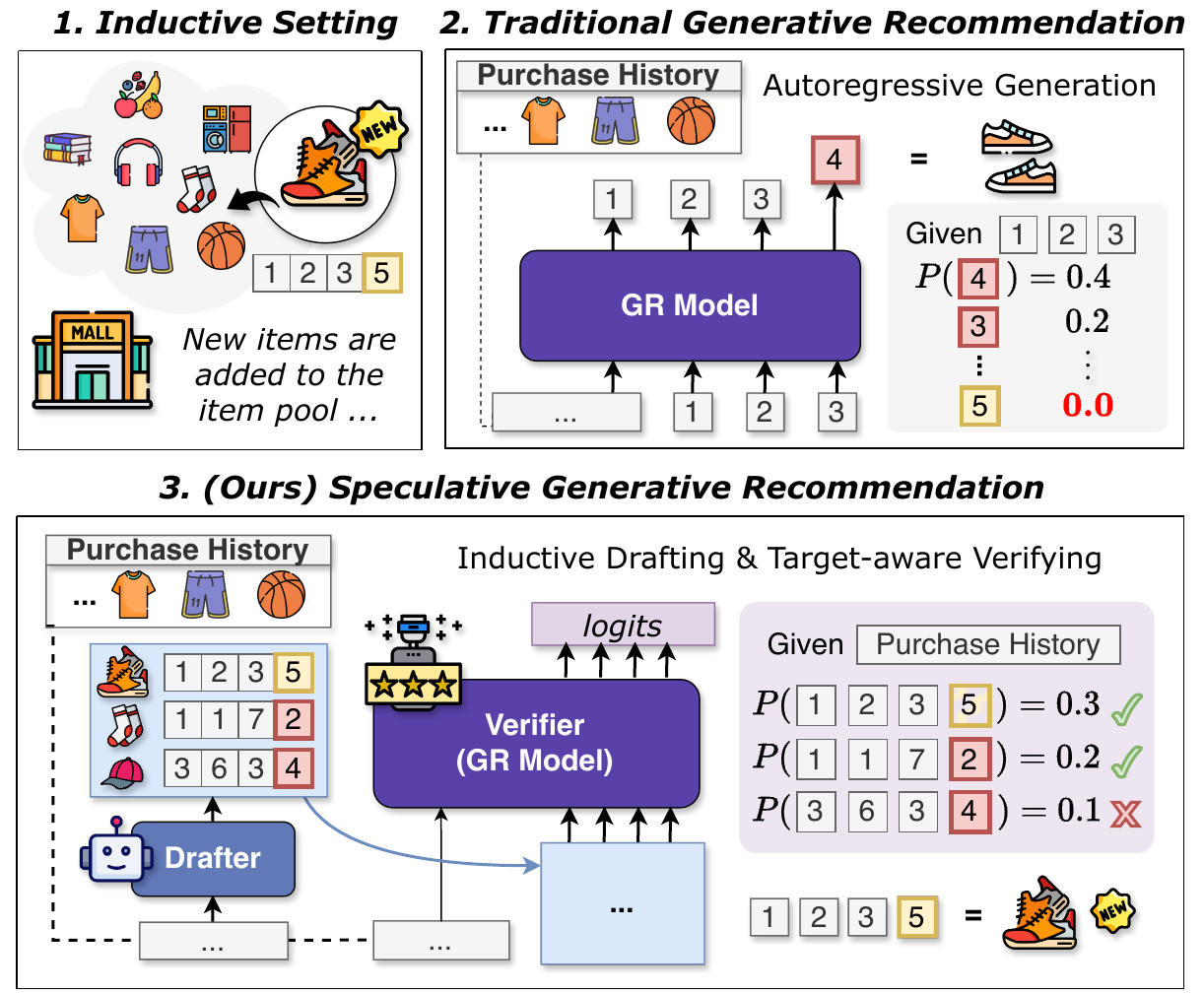}
  \caption{(1 \& 2) GR models struggle to generate unseen items in an inductive setting. (3) SpecGR, a draft-then-verify framework, leverages GR models to verify candidates from an inductive drafter, enabling new-item recommendations.
  }
  \vspace{-0.2in}
  \label{fig:intro}
\end{figure}

Generative recommendation (GR) is an emerging paradigm for sequential recommendation tasks~\cite{rajput2024recommender,zhai2024actions,zheng2024lcrec}. By tokenizing each item into a few discrete tokens (named semantic IDs), models are trained to autoregressively generate the next tokens, which are then parsed as predicted items. Compared to conventional sequential recommendation methods~\cite{kang2018self, sun2019bert4rec}, GR models scale up more easily and achieve better performance~\cite{rajput2024recommender, deng2025onerec}, benefiting from the power of scaling laws~\cite{zhang2023scaling, zhai2024actions}.

Unlike item ID-based transductive models such as SASRec~\cite{kang2018self}, which cannot recommend new items due to the absence of their IDs in the trained model, GR models are expected to exhibit inductive capabilities by capturing semantic correlations and generating semantic tokens for previously unseen items. However, our empirical analysis shows that GR models systematically assign higher likelihoods to seen semantic ID sequences than to unseen ones. As shown quantitatively in~\Cref{tab:main-breakdown}, GR yields near-zero recommendation performance on unseen items. We attribute this to the models' tendency to overfit to the semantic ID patterns present in the training data~\cite{yang2024unifying}, making it unlikely for their outputs to match the semantic IDs of new items (\Cref{fig:intro}). In scenarios where up-to-date recommendations are critical (\eg news or short-video platforms), a flexible, on-the-fly inference framework is essential for the practical deployment of GR models.

In this work, we aim to develop \emph{inductive} generative recommendation models that can recommend new items on-the-fly. Achieving this goal is non-trivial. In traditional recommendation systems, inductive recommendation~\cite{wu2021towards} is typically achieved by incorporating side information and K-nearest neighbor (KNN) search. 
Although GR models can also tokenize items with side information into semantic IDs, as previously discussed, they struggle to generate unseen semantic ID patterns through autoregressive decoding~\cite{freitag2017beam,rajput2024recommender}.
Recent efforts have explored blending GR outputs with items retrieved by non-GR methods~\cite{rajput2024recommender,yang2024unifying}. However, these approaches fail to fully exploit the modeling strengths of GR models, resulting in suboptimal performance.

To this end, we propose \textbf{SpecGR} (\textbf{Spec}ulative \textbf{G}enerative \textbf{R}ecommendation), an inductive generative recommendation framework that can be integrated with GR models in a plug-and-play manner.
We extend the concept of the drafter-verifier framework in the original speculative decoding technique~\cite{leviathan2023SpecDec, chen2023accelerating, he2023rest}.
Rather than relying on a lightweight homologous model for inference acceleration, we explore the integration of models with different paradigms and capabilities. Specifically, we employ a KNN-based inductive model as the drafter to generate small batches of candidate items, while a GR model with stronger recommendation capabilities serves as the verifier, responsible for accepting or rejecting these candidates. Instead of merely blending outputs from different models, SpecGR ensures that all final recommendations are ranked based on scores assigned by the GR verifier.
In addition, we propose guided re-drafting to improve the quality of candidate items proposed by the drafter model, leveraging the semantic ID prefixes generated by the verifier GR model.
Furthermore, to reduce the overhead of maintaining a separate drafter model, we introduce \textbf{SpecGR++} that enables the encoder of the GR model to serve as a drafter.

Extensive experiments are conducted on three public datasets. We split the training and evaluation sets chronologically using fixed timestamp cut-offs. This setup ensures that recommendation models are evaluated in a setting
where new items appear over time. 
The experimental results demonstrate that SpecGR significantly improves the ability of GR models to recommend new items and achieves strong overall performance compared to existing methods.

\section{Related Work}
\label{sec:related-work}

\paratitle{Generative recommendation.} 
\label{sec:related-work-gen-rec}
Traditional sequential recommendation typically assigns a unique learnable embedding to each item, leading to optimization challenges due to large item vocabulary size~\cite{hidasi2016gru4rec,kang2018self,sun2019bert4rec}. Generative recommendation (GR) addresses this by tokenizing items into discrete tokens and predicting the next item via autoregressive next-token generation~\cite{rajput2024recommender,zheng2024lcrec,liu2024multi,hou2025actionpiece}. GR has demonstrated improved memory efficiency~\cite{rajput2024recommender,hou2025generating}, scalability~\cite{zhai2024actions}, and promising end-to-end retrieval capabilities to unify retrieval and ranking~\cite{deng2025onerec}. Despite these advantages, GR models tend to generate only semantic IDs observed during training, severely limiting their generalizability to unseen items. Despite a few early attempts~\cite{rajput2024recommender, yang2024unifying}, this direction remains largely underexplored. In this work, we focus on developing effective frameworks that extend GR models to inductively recommend new items.

\paratitle{Cold-start \& inductive recommendation.}
\label{sec:related-work-inductive-rec}
The item cold-start problem refers to the challenge of recommending items with limited interactions~\cite{zhang2025cold,wei2021contrastive,zhou2023contrastive}.
Common approaches use meta-learning that helps models generalize better from limited interactions,
via gradient-based optimization~\cite{lee2019melu}, task adaptation~\cite{lin2021task,wu2023m2eu}, or memory-augmented modules~\cite{dong2020mamo,zheng2021cold}.
Inductive recommendation particularly tackles the challenge of recommending new items without any associated interactions. Existing methods leverage side information like tags or descriptions~\cite{pazzani2007content,zhu2020recommendation,li2023text}, modality representations~\cite{hou2022towards,hou2023vqrec,sheng2024language}, and behavior patterns~\cite{wu2021towards,wu2020learning}.
However, for GR models, the unique challenge lies in their intrinsic inability to generate unseen items during autoregressive token generation. Recent approaches enhance inductive capability by integrating heuristic candidates~\cite{rajput2024recommender} or dense retrieval results~\cite{yang2024unifying} with GR outputs. In this work, we propose a novel framework that converts the role of GR from generation to verification for inductive recommendation.

\begin{figure*}[t]
  \centering
  \includegraphics[width=\textwidth]{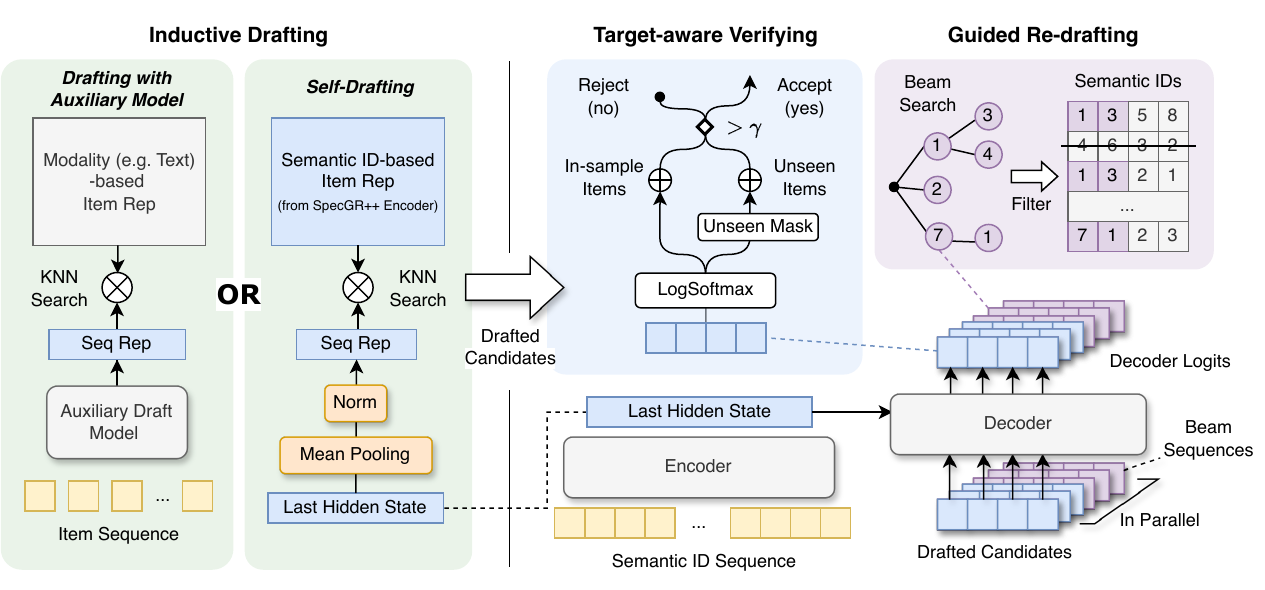}
  \caption{Illustration of the proposed SpecGR method, a draft-and-verify framework that iteratively performs drafting and verification until enough items are accepted. (a) \textbf{Inductive Drafting}. The inductive drafter first retrieves a set of candidates that contain new items. We present two drafting methods: using an auxiliary model or the GR’s encoder output (namely, self-drafting) for item retrieval. (b) \textbf{Target-aware Verifying}. The GR model accepts or rejects the candidates based on the likelihood of being the target. (c) \textbf{Guided Redrafting}. If not enough items are accepted, the GR filters the candidate space for the next drafting round based on the generated beam sequences.}
  \label{fig:main-architecture}
\end{figure*}

\paratitle{Speculative decoding.}
\label{sec:related-work-spec-dec}
Large language models (LLMs)~\cite{achiam2023gpt,zhao2023survey} have revolutionized
a wide range of applications, yet suffer from high inference latency due to autoregressive decoding and large model size. Speculative decoding was proposed to accelerate LLM inference by leveraging a lightweight, homologous drafter model to generate future token sequences, which are then verified by a stronger target model,
achieving lower latency and lossless performance~\cite{leviathan2023SpecDec, chen2023accelerating}. Subsequent efforts have developed efficient drafting strategies~\cite{cai2024medusa, he2023rest, elhoushi2024layer}, improved draft quality to maximize acceptance rate~\cite{xiao2024clover, gloeckle2024better}, and applied the framework to diverse domains~\cite{wang2024speculative, de2025accelerated}.
Recently, ~\citet{lin2024efficient} extended speculative decoding to accelerate top-K sequence generation in the GR setting. In this work, rather than focusing on acceleration, we leverage the draft-verify framework to bring inductive capabilities to generative recommendation models.

\section{Methodology}
\label{sec:methodology}

\subsection{Problem Setup and Formulation}
\label{subsec:problem-setup-and-formulation}

We follow the inductive sequential recommendation task~\cite{hou2022towards,yuan2023go}. The input is a sequence of items $\{x_1, x_2, \dots, x_w\}$ ordered chronologically based on user interaction time. Each item $x \in \mathcal{I}$ has associated text features, such as title, description, and category. Here, $w$ denotes the length of the item sequence. The task is to predict the next item of interest. Note that in the inductive setting, the target item may not appear in the training set, called new or unseen items in the following sections.

In this work, we focus on developing a general inductive recommendation framework for generative recommendation models.
Typical GR models like TIGER~\cite{rajput2024recommender} will tokenize each item $x_i$ into a semantic ID pattern, $\text{ID}_{i} := [ \langle c_1^i \rangle, \langle c_2^i \rangle, \ldots, \langle c_{l}^i \rangle ]$, where $l$ denotes the number of digits of one item's semantic ID pattern, and $\langle c \rangle$ denotes one digit of semantic ID.
In this way, the input for GR models can be represented as follows by replacing the items in the original item sequence with the corresponding semantic IDs:
\begin{equation}
X = \left[\langle\texttt{bos} \rangle, \text{ID}_1, \text{ID}_2, \ldots, \text{ID}_w, \langle \texttt{eos} \rangle\right],\label{eqn:input}
\end{equation}
where $\langle\texttt{bos} \rangle$ and $\langle \texttt{eos} \rangle$ are special tokens indicating the start and ending positions of a semantic ID sequence. Then the GR models are trained to generate $K$ semantic ID patterns, which will be further parsed into recommended items with top-$K$ probabilities.
In the inductive setting, we assume that new items have been assigned semantic ID patterns. These new items can be parsed if the outputs of the GR models match the new semantic ID patterns.

\subsection{Speculative Generative Recommendation}
\label{subsec:speculative-generative-recommendation}

We begin by providing an overview of the proposed SpecGR framework, as illustrated in~\Cref{fig:main-architecture}. The framework consists of four components:
(1) \textbf{Inductive Drafting.}
Given the same input item sequence as the generative recommendation model, a drafter model with inductive recommendation capabilities proposes a set of candidate items as recommendation drafts.
(2) \textbf{Target-aware Verifying.}
The GR model, acting as a verifier, either accepts or rejects the candidates based on the probability that they could be targets of the input sequence.
(3) \textbf{Guided Re-drafting.}
If the number of accepted items does not meet the required recommendations $K$, the GR model guides the drafter to re-draft and propose the next set of candidate items.
(4) \textbf{Adaptive Exiting.}
Once $K$ items are accepted, the framework exits and outputs the items, sorted by the scores given by the GR model.
This process selects high-quality, unseen items through drafting and verifying, while maintaining the strong recommendation capabilities of generative recommendation models.

\subsubsection{Inductive Drafting}\label{subsubsec:ind_drafting} Instead of expecting GR models to directly generate the semantic IDs of unseen items, we employ an inductive drafter model to first propose ``recommendation drafts'' that may include unseen items. Given the input item sequence $X$, the drafter model $\operatorname{D}(\cdot)$ performs inductive drafting by recommending a set of $\delta$ candidates
$\mathcal{Q} = \operatorname{D}(X)$, where $|\mathcal{Q}| = \delta$. 
Note that in the original speculative decoding technique, the drafter model is considered an efficient approximation of the target model~\cite{leviathan2023SpecDec, chen2023accelerating}, functioning as a homologous model to the verifier model. Here, we extend the above restriction in choosing drafter models. We do not require the drafter model to be a GR model but instead an inductive model to bring new capabilities to the following verifier model (in our case, the GR model).
This approach allows high-quality unseen items to be introduced into the system, which are then verified by the more expressive generative model and finally included in the recommendations. Hence, the inductive drafter can be any inductive recommendation model, and we will discuss two effective instantiations in the subsequent Drafting Strategies section.

\subsubsection{Target-aware Verifying}\label{subsubsec:target_aware_verify}

While inductive drafters excel at recommending unseen items, they are not as effective as GR models in modeling input sequences and providing recommendations. Therefore, after obtaining candidates $\mathcal{Q}$ from the inductive drafting process, we use the generative model to verify them, rejecting items with low likelihood.

$\bullet$ \textit{\textbf{Target-aware likelihood for ranking.}}
Given a candidate item as a potential target, we use the GR model as a query-likelihood model (QLM)~\cite{zhuang2023open, zhuang2021deep, nogueira2019doc2query}.
The QLM scores the query, \ie input sequence and the potential target, by measuring the likelihood of the model consecutively generating the tokens in the query.
Previous studies have demonstrated that generative language models, such as T5~\cite{raffel2020exploring}, exhibit robust zero-shot query-likelihood ranking performance in document retrieval tasks, without explicit fine-tuning for document ranking~\cite{zhuang2023open}. Accordingly, conditioning on the input sequence $X$, we adopt the conditional probability of generating the target semantic ID pattern as the verification score.

$\bullet$ \textit{\textbf{Likelihood score calculation.}} However, naively applying the QLM for verification would result in low scores for unseen items. This happens because not all digits of semantic IDs are derived from item semantics.
In addition to the tokens learned purely from item semantics, existing methods usually add an extra digit to avoid conflicts, known as the item identification token~\cite{rajput2024recommender,liu2024multi}. For unseen items, the probability of generating this identification token lies outside the modeling distribution and therefore primarily consists of noise.

To provide a fair verification score for unseen items, we exclude the identification token and calculate only the probability of other digits. The target-aware verification score can be computed as:
\begin{equation}
\label{equation:qlm-score}
\operatorname{V}(x_t, X) = 
\begin{cases} 
\displaystyle \frac{1}{l} \sum_{i=1}^{l} \log P(c^t_i \mid c^t_{<i}, X) & \text{if } x_t \in \mathcal{I}, \\[15pt]
\displaystyle \frac{1}{l-1} \sum_{i=1}^{l-1} \log P(c^t_i \mid c^t_{<i}, X) & \text{if } x_t \in \mathcal{I}^* \setminus \mathcal{I} ,
\end{cases}
\end{equation}
where $\operatorname{V}(\cdot)$ denotes the verifier model, which takes the input sequence $X$ and the potential target item $x_t$ as inputs, outputing the log-likelihood probability scores. $l$ denotes the total number of digits in each semantic ID pattern, where the last digit is assumed to be the item identification token. $P(\cdot)$ denotes the backbone autoregressive model, which takes semantic ID sequences and outputs the likelihood scores. $c^t_i$ refers to the $i$-th digit of the semantic ID for the target item $x_t$. The set $\mathcal{I}^* \setminus \mathcal{I}$ represents the unseen items.

To alleviate the bias caused by varying lengths of semantic IDs for unseen and existing items, we normalize the log-likelihood scores by the corresponding number of digits. After obtaining the likelihood score, we accept the items if \(\operatorname{V}(x_t,X) > \gamma\), where \(\gamma\) is a hyperparameter and can be tuned on the validation set.

\subsubsection{Guided Re-drafting}\label{subsubsec:guided_re_draft}

If fewer than $K$ items are accepted from the initial batch of drafted candidates, the drafter model $\operatorname{D}(\cdot)$ must generate an additional batch of $\delta$ new candidates. Intuitively, since these candidates appear lower in the drafter’s ranking, their expected acceptance probability under the verifier $\operatorname{V}(\cdot)$ is substantially reduced. To counteract this decline in acceptance rate, we introduce the \emph{guided re-drafting} mechanism that steers subsequent candidate batches toward regions of the item space that better match the verifier’s scoring distribution.

Guided re-drafting operates by steering the drafter models using a set of semantic ID prefixes generated by the verifier models (GR models). Specifically, after verifying the $j$-th batch of recommendation drafts, the verifier model generates a set of beam sequences $\mathcal{B}_j$ using beam search, where each sequence is a $j$-digit semantic ID prefix. In the next draft-verify iteration, the drafter model is guided to propose only candidates $\mathcal{Q}_j$ whose prefixes match those in $\mathcal{B}_j$:
\begin{equation}
\mathcal{Q}_j = \left\{x_i | x_i \in \operatorname{D}(X), (c_1^i, c_2^i, \ldots, c_j^i) \in \mathcal{B}_j \right\},
\end{equation}
where $\mathcal{B}_j$ denotes the set of semantic ID prefixes, with a hyperparameter $\beta$ as the set size.
Guided re-drafting happens along with the beam search decoding process of the GR model. It is important to note that the total number of draft-verify iterations will not exceed $l$, which corresponds to the maximum length of the semantic IDs, and is also equal to the maximum number of decoding steps.

\subsubsection{Adaptive Exiting}\label{subsubsec:ada_exit}

SpecGR can adaptively terminate the draft-verify iterations based on the number of candidate items accepted by the verifier (GR) model. When the number of accepted items reaches \(K\), the loop exits, avoiding the need to generate full-length sequences of \(l\).
This adaptive approach reduces inference time, as fewer generation steps are required.
If, after the final iteration, there are still not enough accepted items, the beam sequences will be appended to the recommendation list until \(K\) is reached. As a result, even in the worst case, SpecGR does not incur additional time overhead compared to decoding with beam search. Finally, we rank the recommendation list by using the verification scores of the accepted items, along with the beam scores if items from beam sequences are included.

\subsection{Drafting Strategies}
\label{subsec:drafting-strategies}

In this section, we present two methods for drafting: by using an auxiliary draft model, and by reusing the encoder of generative recommendation models (namely SpecGR++).

\subsubsection{Auxiliary Draft Model}\label{subsubsec:draft-aux-model}

The most straightforward way to draft is to introduce an auxiliary inductive recommendation model. An example is UniSRec~\cite{hou2022towards}, which uses modality-based item representations for KNN search. When new items are added, their representations can be directly incorporated into the item pool. The model can then retrieve new items if their modality-based representations are similar to the sequence representations.

\subsubsection{Self-Drafting via GR Encoder}\label{subsubsec:self-spec}

Despite the flexibility of using an auxiliary model as the drafter, issues such as communication latency and distribution shift may arise. Thus, we propose \textbf{SpecGR++}, which reuses the encoder module of the generative recommendation model to function as an inductive drafter model.
The general idea is to encode both (1) the semantic IDs of a single item, and (2) the input semantic ID sequence of user history, using the same encoder module. Then we apply KNN search to retrieve semantic IDs of both existing and new items.

$\bullet$ \textit{\textbf{Semantic ID-based item and sequence representations.}}
To derive sequence representations, we use the same input format as our GR model (\Cref{eqn:input}). 
To derive item representations, we format the semantic IDs of a single item $x_i$ in the same way as the encoder's input, \ie $[\langle \texttt{bos} \rangle, \text{ID}_{i}, \langle \texttt{eos} \rangle]$, where \(\text{ID}_{i}\) represents the semantic ID pattern of item $x_i$.
To obtain the item and sequence representations, we take the last hidden state from the GR encoder and apply mean pooling.

$\bullet$ \textit{\textbf{Item–sequence contrastive pretraining.}}
Prior work shows that hidden states from generative models are not directly suitable as representations~\cite{ni2021sentence,ni2021large}. 
To obtain strong inductive embeddings, we jointly train the GR encoder with a contrastive objective following~\cite{chen2020simple,hou2022towards}. 
Given sequence embeddings $\bm{s}_j$ and next-item embeddings $\bm{i}_j$, the contrastive loss is:
\begin{equation}
\mathcal{L}_{\text{CL}}
    = -\frac{1}{B_{\text{emb}}}
    \sum_{j=1}^{B_{\text{emb}}}
    \log
    \frac{
        \exp(\bm{s}_j \cdot \bm{i}_j / \tau)
    }{
        \sum_{j'=1}^{B_{\text{emb}}}
        \exp(\bm{s}_j \cdot \bm{i}_{j'} / \tau)
    }.
\label{eq:loss_cl}
\end{equation}
We optimize $\mathcal{L}_{\text{CL}}$ jointly with the next-token generation loss:
\begin{equation}
\mathcal{L}_{\text{Gen}}
= -\frac{1}{B_{\text{gen}}}
  \sum_{b=1}^{B_{\text{gen}}}
  \frac{1}{L_b}
  \sum_{t=1}^{L_b}
  \log P(c_{b,t} \mid c_{b,<t}, X_b).
\label{eq:loss_gen}
\end{equation}
To ensure sufficient in-batch negatives for the contrastive objective, we use a larger embedding batch (\ie $B_{\text{emb}} > B_{\text{gen}}$). The final multi-task loss is expressed as $\mathcal{L}
= \lambda_1 \mathcal{L}_{\text{CL}}
+ \mathcal{L}_{\text{Gen}}$, where $\lambda_1$ is a hyperparameter balancing the two tasks.

$\bullet$ \textit{\textbf{Learning-to-rank fine-tuning.}}
Following~\citet{li2023text}, to further enhance the ranking ability of the semantic ID encoder, we continue to fine-tune the encoder using the cross-entropy loss $\mathcal{L}_{\text{CE}}$ on a larger batch of negative items (equivalent to $\mathcal{L}_{\text{CL}}$ in~\Cref{eq:loss_cl} with $B_{\text{emb}} = |\mathcal{I}|$).
To enable efficient large-batch training, the item representations are frozen at the beginning of the fine-tuning phase.
The overall loss for fine-tuning can be written as $\mathcal{L}' = \lambda_2\mathcal{L}_{\text{CE}} + \mathcal{L}_{\text{Gen}}$, where $\lambda_2$ is a hyperparameter.

\definecolor{highlight_bg_color}{rgb}{0.95, 0.95, 0.95} %

\begin{table*}[!ht]
\large
\centering
\caption{Performance comparison of different models. The best and the second-best performance is denoted in bold and underlined fonts, respectively. ``R@K'' is short for ``Recall@K'' and ``N@K'' is short for ``NDCG@K''. 
``Improv.'' denotes the improvement ratio of SpecGR compared to the best-performing baseline model. }
\label{tab:main-overall}
\resizebox{\textwidth}{!}{
\begin{tabular}{@{}cccccccccccccc@{}}
\toprule
\multirow{2.5}{*}{\textbf{Dataset}} & \multirow{2.5}{*}{\textbf{Metric}}
& \multicolumn{1}{c}{\textbf{ID-based}} 
& \multicolumn{2}{c}{\textbf{Feature + ID}} 
& \multicolumn{3}{c}{\textbf{Modality-based}} 
& \multicolumn{3}{c}{\textbf{Generative}} 
& \multicolumn{2}{c}{\cellcolor{highlight_bg_color}\textbf{Ours}} 
& \multirow{2.5}{*}{\textbf{Improv.}} \\ 
\cmidrule(l){3-3} \cmidrule(l){4-5} \cmidrule(l){6-8} \cmidrule(l){9-11} \cmidrule(l){12-13}
&  & \textbf{SASRec}\textsubscript{ID} & \textbf{FDSA} & \textbf{S$^3$-Rec} & \textbf{SASRec}\textsubscript{T} & \textbf{UniSRec} & \textbf{Recformer} & \textbf{TIGER} & \textbf{TIGER}\textsubscript{C} & \textbf{LIGER} & \cellcolor{highlight_bg_color}\textbf{SpecGR}\textsubscript{Aux} & \cellcolor{highlight_bg_color}\textbf{SpecGR++} &  \\ 
\midrule
\multirow{4}{*}{Games} &
R@10 & 0.0186 & 0.0190 & 0.0195 & 0.0179 & 0.0225 & 0.0243 & 0.0222 & 0.0226 & 0.0139 & \cellcolor{highlight_bg_color}\textbf{0.0254} & \cellcolor{highlight_bg_color}\underline{0.0250} & +4.53\% \\
& N@10   & 0.0093 & 0.0101 & 0.0094 & 0.0091 & 0.0115 & 0.0111 & 0.0114 & 0.0115 & 0.0068 & \cellcolor{highlight_bg_color}\textbf{0.0128} & \cellcolor{highlight_bg_color}\underline{0.0124} & +10.40\% \\
& R@50 & 0.0477 & 0.0496 & 0.0473 & 0.0507 & 0.0621 & \underline{0.0740} & 0.0584 & 0.0611 & 0.0635 & \cellcolor{highlight_bg_color}\textbf{0.0778} & \cellcolor{highlight_bg_color}0.0717 & +5.13\% \\
& N@50   & 0.0162 & 0.0167 & 0.0154 & 0.0161 & 0.0200 & 0.0218 & 0.0193 & 0.0198 & 0.0172 & \cellcolor{highlight_bg_color}\textbf{0.0239} & \cellcolor{highlight_bg_color}\underline{0.0225} & +9.72\% \\
\midrule
\multirow{4}{*}{Office} &
R@10 & 0.0093 & 0.0095 & 0.0100 & 0.0091 & 0.0119 & 0.0126 & 0.0132 & 0.0130 & 0.0059 & \cellcolor{highlight_bg_color}\textbf{0.0138} & \cellcolor{highlight_bg_color}\underline{0.0134} & +3.99\% \\
& N@10   & 0.0047 & 0.0050 & 0.0052 & 0.0048 & 0.0062 & 0.0039 & \underline{0.0071} & 0.0070 & 0.0029 & \cellcolor{highlight_bg_color}\textbf{0.0072} & \cellcolor{highlight_bg_color} 0.0070 & +1.68\% \\
& R@50 & 0.0217 & 0.0224 & 0.0234 & 0.0233 & 0.0322 & \underline{0.0340} & 0.0308 & 0.0312 & 0.0268 & \cellcolor{highlight_bg_color}\textbf{0.0360} & \cellcolor{highlight_bg_color}0.0332 & +5.93\% \\
& N@50   & 0.0074 & 0.0078 & 0.0080 & 0.0078 & 0.0105 & 0.0106 & 0.0109 & 0.0110 & 0.0072 & \cellcolor{highlight_bg_color}\textbf{0.0119} & \cellcolor{highlight_bg_color}\underline{0.0113} & +8.76\% \\
\midrule
\multirow{4}{*}{Phones} &
R@10 & 0.0052 & 0.0067 & 0.0058 & 0.0072 & 0.0084 & 0.0074 & 0.0090 & 0.0087 & 0.0048 & \cellcolor{highlight_bg_color}\underline{0.0099} & \cellcolor{highlight_bg_color}\textbf{0.0101} & +11.90\% \\
& N@10   & 0.0027 & 0.0035 & 0.0028 & 0.0037 & 0.0045 & 0.0036 & 0.0047 & 0.0046 & 0.0022 & \cellcolor{highlight_bg_color}\underline{0.0050} & \cellcolor{highlight_bg_color}\textbf{0.0052} & +10.64\% \\
& R@50 & 0.0143 & 0.0184 & 0.0151 & 0.0188 & 0.0233 & 0.0236 & 0.0232 & 0.0233 & 0.0226 & \cellcolor{highlight_bg_color}\textbf{0.0285} & \cellcolor{highlight_bg_color}\underline{0.0275} & +20.64\% \\
& N@50   & 0.0047 & 0.0060 & 0.0048 & 0.0062 & 0.0077 & 0.0070 & 0.0078 & 0.0078 & 0.0059 & \cellcolor{highlight_bg_color}\textbf{0.0090} & \cellcolor{highlight_bg_color}\textbf{0.0090} & +14.80\% \\
\bottomrule
\end{tabular}
}
\end{table*}

\begin{table*}[t]
\large
\centering
\caption{Model performance breakdown on the ``in-sample'' and ``unseen'' subsets. The proportions of the test cases in each subset relative to the entire test data have been labeled. The best and second-best results are bolded and underlined.}
\label{tab:main-breakdown}
\resizebox{\textwidth}{!}{
\setlength{\tabcolsep}{2.0mm}%
\begin{tabular}{@{\hspace{0.6mm}}ccccccccccccccc@{}}
\toprule
\multirow{3}{*}{\textbf{Model}} & \multirow{3}{*}{
\makecell{\\ \textbf{\#Params.} \\ \textbf{(M)}}
} & \multicolumn{6}{c}{\textbf{Games}} & \multicolumn{6}{c}{\textbf{Phones}} \\ \cmidrule(l){3-8} \cmidrule(l){9-14}
 &  & \multicolumn{2}{c}{\cellcolor{highlight_bg_color}\textbf{Overall}} & \multicolumn{2}{c}{\makecell{\textbf{In-Sample} \\ \footnotesize{(39.7\%)}}} & \multicolumn{2}{c}{\makecell{\textbf{Unseen} \\ \footnotesize{(60.3\%)}}} & \multicolumn{2}{c}{\cellcolor{highlight_bg_color}\textbf{Overall}} & \multicolumn{2}{c}{\makecell{\textbf{In-Sample} \\ \footnotesize{(31.8\%)}}} & \multicolumn{2}{c}{\makecell{\textbf{Unseen} \\ \footnotesize{(68.2\%)}}} \\ \cmidrule(l){3-4} \cmidrule(l){5-6} \cmidrule(l){7-8} \cmidrule(l){9-10} \cmidrule(l){11-12} \cmidrule(l){13-14}
 & & \cellcolor{highlight_bg_color}\textbf{R@50} & \cellcolor{highlight_bg_color}\textbf{N@50} & \textbf{R@50} & \textbf{N@50} & \textbf{R@50} & \textbf{N@50} & \cellcolor{highlight_bg_color}\textbf{R@50} & \cellcolor{highlight_bg_color}\textbf{N@50} & \textbf{R@50} & \textbf{N@50} & \textbf{R@50} & \textbf{N@50} \\ \midrule
UniSRec  & 2.90  & \cellcolor{highlight_bg_color}0.0621 & \cellcolor{highlight_bg_color}0.0200 & 0.1386 & 0.0461 & 0.0118 & 0.0029 & \cellcolor{highlight_bg_color}0.0233 & \cellcolor{highlight_bg_color}0.0077 & 0.0604 & 0.0211 & 0.0060 & 0.0014 \\
Recformer  & 233.73  & \cellcolor{highlight_bg_color}\underline{0.0740} & \cellcolor{highlight_bg_color}0.0218 & 0.1082 & 0.0333 & \underline{0.0514} & \underline{0.0142} & \cellcolor{highlight_bg_color}0.0236 & \cellcolor{highlight_bg_color}0.0070 & 0.0340 & 0.0103 & \textbf{0.0188} & \textbf{0.0055} \\
TIGER  & 13.26  & \cellcolor{highlight_bg_color}0.0584 & \cellcolor{highlight_bg_color}0.0193 & \underline{0.1472} & \textbf{0.0486} & - & - & \cellcolor{highlight_bg_color}0.0232 & \cellcolor{highlight_bg_color}0.0078 & \underline{0.0730} & \underline{0.0245} & - & - \\
TIGER\textsubscript{C} & 13.26  & \cellcolor{highlight_bg_color}0.0611 & \cellcolor{highlight_bg_color}0.0198 & 0.1447 & \underline{0.0482} & 0.0061 & 0.0011 & \cellcolor{highlight_bg_color}0.0233 & \cellcolor{highlight_bg_color}0.0078 & 0.0691 & 0.0238 & 0.0019 & 0.0003 \\
LIGER & 13.26  & \cellcolor{highlight_bg_color}0.0635 & \cellcolor{highlight_bg_color}0.0172 & 0.0438 & 0.0160 & \textbf{0.0765} & \textbf{0.0179} & \cellcolor{highlight_bg_color}0.0226 & \cellcolor{highlight_bg_color}0.0059 & 0.0472 & 0.0107 & \underline{0.0111} & \underline{0.0037} \\
\midrule
\textbf{SpecGR\textsubscript{Aux}}  & 16.16  & \cellcolor{highlight_bg_color}\textbf{0.0778} & \cellcolor{highlight_bg_color}\textbf{0.0239} & \textbf{0.1485} & 0.0457 & 0.0312 & 0.0096 & \cellcolor{highlight_bg_color}\textbf{0.0285} & \cellcolor{highlight_bg_color}\textbf{0.0090} & \textbf{0.0748} & 0.0237 & 0.0069 & 0.0021 \\
\textbf{SpecGR++}  & 13.28  & \cellcolor{highlight_bg_color} 0.0717 & \cellcolor{highlight_bg_color} \underline{0.0225} & 0.1323 & 0.0439 & 0.0318 & 0.0084 & \cellcolor{highlight_bg_color}\underline{0.0275} & \cellcolor{highlight_bg_color}\textbf{0.0090} & \underline{0.0730} & \textbf{0.0246} & 0.0063 & 0.0017 \\ \bottomrule
\end{tabular}
}
\end{table*}

\section{Experiments}

In this section, we present experimental results to answer the following Research Questions (\textbf{RQ}).
\begin{itemize}
  \item \textbf{RQ1:} How does SpecGR perform compared with state-of-the-art baselines on overall and seen/unseen subsets?
  \item \textbf{RQ2:} Do all of SpecGR's designs take effect?
  \item \textbf{RQ3:} How do key hyperparameters affect SpecGR's performance and efficiency?
  \item \textbf{RQ4:} Is SpecGR an effective plug-and-play framework for different drafters and GR backbones?
\end{itemize}

\subsection{Experimental Setup}

\subsubsection{Datasets}

We use three categories, Video Games (\textbf{Games}), Office Products (\textbf{Office}), and Cell Phones and Accessories (\textbf{Phones}), from the Amazon Reviews 2023 dataset~\cite{hou2024bridging} as our experimental datasets.
To assess the performance of SpecGR in real-world settings, we
utilize preprocessed benchmarks\footnote{https://huggingface.co/datasets/McAuley-Lab/Amazon-Reviews-2023/tree/main/benchmark/5core/timestamp\_w\_his} that exclude users and items with fewer than five interactions. The data is split into training, validation, and test sets based on predefined timestamp cut-offs. 
Notably, since the datasets are split by timestamps, the validation and test sets naturally include unseen items. This simulates a more realistic scenario in comparison to the widely-used leave-last-out splitting.

\begin{table*}[t]
\centering
\caption{Ablation study on SpecGR++ inference and training. The best and second-best results are bolded and underlined.}
\label{tab:ablation-main}
\resizebox{\textwidth}{!}{%
\setlength{\tabcolsep}{1.2mm}
\begin{tabular}{lcccccccccccc}
\toprule
\multirow{3}{*}{\textbf{Variants}} & \multicolumn{4}{c}{\textbf{Games}} & \multicolumn{4}{c}{\textbf{Office}} & \multicolumn{4}{c}{\textbf{Phones}} \\ 
\cmidrule(l){2-5}\cmidrule(l){6-9}\cmidrule(l){10-13}
 & \textbf{R@50} & \textbf{N@50} & \textbf{R@10} & \textbf{N@10} & \textbf{R@50} & \textbf{N@50} & \textbf{R@10} & \textbf{N@10} & \textbf{R@50} & \textbf{N@50} & \textbf{R@10} & \textbf{N@10} \\ 
\midrule
(1.1) \emph{w/o} inductive drafting            & 0.0609 & 0.0202 & 0.0235 & \underline{0.0121} & 0.0306 & 0.0109 & \underline{0.0132} & \textbf{0.0070} & 0.0233 & 0.0080 & 0.0092 & 0.0049 \\
(1.2) \emph{w/o} likelihood score adjustment   & \underline{0.0712} & 0.0221 & 0.0236 & 0.0119 & 0.0331 & 0.0103 & 0.0118 & 0.0057 & 0.0236 & 0.0081 & 0.0092 & 0.0049 \\
(1.3) \emph{w/o} guided re-drafting            & 0.0611 & 0.0202 & 0.0235 & \underline{0.0121} & 0.0309 & 0.0110 & \underline{0.0132} & \textbf{0.0070} & 0.0264 & 0.0086 & 0.0096 & 0.0050 \\
(1.4) \emph{w/o} item re-ranking               & 0.0703 & 0.0219 & \underline{0.0239} & 0.0120 & \textbf{0.0334} & \textbf{0.0113} & 0.0131 & 0.0069 & 0.0264 & 0.0083 & 0.0093 & 0.0047 \\
(1.5) \emph{w/o} adaptive exiting              & 0.0694 & 0.0200 & 0.0203 & 0.0095 & 0.0313 & 0.0108 & 0.0126 & 0.0068 & \underline{0.0265} & 0.0086 & 0.0095 & 0.0050 \\ 
\cmidrule(r){1-13}
(2.1) TIGER for SpecGR++                       & 0.0582 & 0.0192 & 0.0224 & 0.0114 & 0.0302 & 0.0105 & 0.0127 & 0.0067 & 0.0232 & 0.0078 & 0.0090 & 0.0047 \\
(2.2) \emph{w/o} contrastive pretraining       & 0.0581 & 0.0193 & 0.0221 & 0.0115 & 0.0313 & 0.0108 & 0.0126 & 0.0068 & 0.0234 & 0.0077 & 0.0093 & 0.0050 \\
(2.3) \emph{w/o} fine-tuning                   & 0.0692 & \textbf{0.0225} & 0.0222 & 0.0111 & 0.0325 & 0.0110 & 0.0129 & 0.0068 & 0.0259 & \underline{0.0087} & \underline{0.0098} & \underline{0.0051} \\ 
\cmidrule(r){1-13}
\textbf{SpecGR++}                              & \textbf{0.0717} & \textbf{0.0225} & \textbf{0.0250} & \textbf{0.0124} & \underline{0.0332} & \textbf{0.0113} & \textbf{0.0134} & \textbf{0.0070} & \textbf{0.0275} & \textbf{0.0090} & \textbf{0.0101} & \textbf{0.0052} \\
\bottomrule
\end{tabular}%
}
\end{table*}

\subsubsection{Compared Methods}

We report results for two SpecGR variants: SpecGR\textsubscript{Aux}, which uses UniSRec~\cite{hou2022towards} as an auxiliary drafter, and SpecGR++, which uses its own encoder module for drafting. We compare SpecGR against the following state-of-the-art methods: ID-based methods such as SASRec~\cite{kang2018self}; feature+ID-based methods such as FDSA~\cite{zhang2019feature} and S$^3$-Rec~\cite{zhou2020s3}; modality-based methods such as UniSRec~\cite{hou2022towards} and RecFormer~\cite{li2023text}; and generative methods including TIGER, TIGER\textsubscript{C}~\cite{rajput2024recommender}, and LIGER~\cite{yang2024unifying}. Notably, TIGER\textsubscript{C} employs a heuristic strategy that mixes a fixed proportion of unseen items into TIGER’s recommendation list.

\subsubsection{Evaluation Setting}

We adopt Recall@$K$ and NDCG@$K$ as metrics to evaluate the compared methods, where $K \in \{10, 50\}$.
In addition, based on whether the target items in the test set are existing items or new items (not shown in the training set), we split our test set into two subsets, named \textbf{In-Sample} and \textbf{Unseen}, respectively. 
The model checkpoints of all compared methods that have the best overall performance on the validation set will be evaluated on the test set.

\subsubsection{Implementation Details}
\label{sec:experiment-implementation-details}

We use UniSRec~\cite{hou2022towards} as the auxiliary drafter model for SpecGR\textsubscript{Aux} and TIGER~\cite{rajput2024recommender} as the GR backbone for both SpecGR variants. Input sequences are truncated to a maximum of 20 items, following~\citet{rajput2024recommender}, and the same semantic ID tokenization process is applied. SpecGR++ is trained using a multi-task setup (\(\lambda_1 = \lambda_2 = 6.0\)) combining generation and contrastive objectives, followed by a learning-to-rank fine-tuning phase.  We use \(B_{\text{Emb}} = 2048\) and \(B_{\text{Gen}} = 256\). Hyperparameters such as the draft threshold (\(\gamma\)), beam size (\(\beta\)), and draft size (\(\delta\)) are tuned on validation splits.

\subsection{Performance Analysis (RQ1)}\label{sec:main_exp}

\paratitle{Overall performance.} We compare SpecGR with sequential and generative recommendation baselines across three public datasets; results are summarized in~\Cref{tab:main-overall}. ID-based and feature-based methods generally show poor performance, especially on sparse datasets (\eg Phones). Modality-based methods, such as UniSRec and Recformer, show improved performance by leveraging powerful text embeddings from Pretrained Language Models (PLMs). Generative recommendation models achieve the best results through autoregressive modeling of fine-grained semantic IDs. Among all models, SpecGR consistently achieves the best overall performance (\eg up to +14.8\% in NDCG@50, and +20.64\% in Recall@50). Notably, SpecGR++ attains both better parameter efficiency and comparable performance to SpecGR\textsubscript{Aux}, highlighting the GR encoder’s effectiveness in learning robust semantic ID-based representations for inductive recommendation.

\paratitle{Subset Analysis.} Next, we analyze the detailed performance breakdown on the in-sample and unseen subsets, as shown in~\Cref{tab:main-breakdown}. TIGER achieves strong in-sample performance but fails to generalize inductively on the unseen subset. SpecGR addresses this limitation by integrating inductive drafting with GR-based verification, greatly improving inductive generalization without compromising in-sample quality. 
Recformer is the best-performing modality-based method. However, its LLM backbone has significantly larger model size compared to other baselines.
While LIGER attains higher performance on unseen items via dense retrieval-based candidate blending, its heuristic combination of candidates introduces irrelevant items, yielding a suboptimal trade-off and degraded overall recommendation quality. In contrast, SpecGR employs GR’s target-aware likelihood scores to filter inductive candidates, resulting in strong inductive ability and the best overall performance.

\subsection{Ablation Study (RQ2)}
\label{sec:experiments-ablation}

\paratitle{SpecGR Inference Framework.}
First, we assess the contributions of inference components in SpecGR++, as summarized in~\Cref{tab:ablation-main}. Inductive drafting (1.1) and likelihood score adjustment (1.2) substantially enhance inductive generalization by incorporating meaningful unseen candidates. Guided re-drafting (1.3) boosts recommendation quality when initial drafting provides insufficient accepted candidates. Removing verification-based re-ranking (1.4) and adaptive exiting (1.5) results in performance degradation, confirming the importance of these modules in 
maintaining a balance between in-sample accuracy and inductive capability.

\paratitle{SpecGR++ Training Paradigm.}
Next, we analyze training paradigm variants for SpecGR++. Directly using TIGER’s encoder states without dedicated representation learning (2.1) severely limits inductive recommendation. Contrastive pretraining (2.2) significantly improves representation quality, and subsequent fine-tuning (2.3) further refines performance, validating our two-stage training design.

\begin{figure}[t]
  \centering
  \includegraphics[width=1.02\linewidth, clip, trim=0 0 0 0]{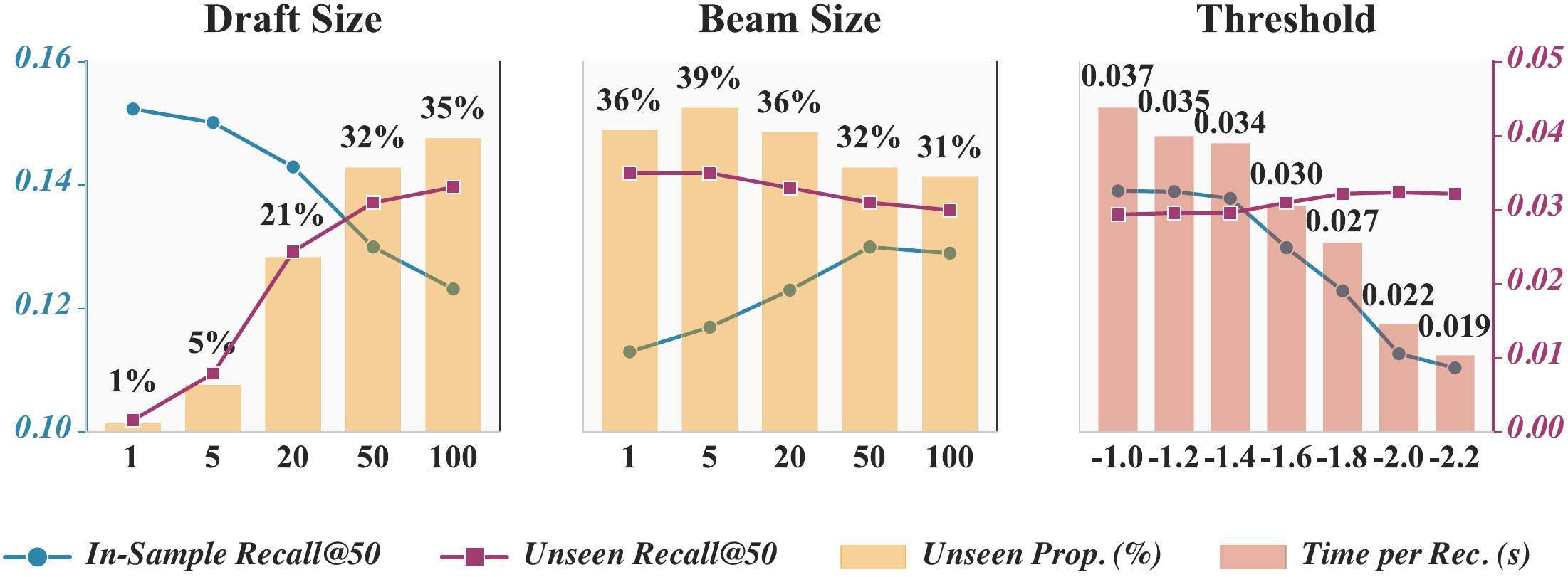}
    \caption{Impact of hyperparameters on SpecGR's performance and efficiency. (Left, middle): Bars show the proportion of unseen items in recommendations. (Right): Bars represent inference latency in seconds. Lines depict the trade-off between in-sample and unseen Recall@50.}
    
  \label{fig:hyper_combined}
\end{figure}

\subsection{Hyperparameter Analysis (RQ3)}
\label{sec:experiments-hyperparameters}
We analyze how model hyperparameters affect recommendation behaviors and performance. We conduct hyperparameter analyses of \textbf{SpecGR++} on the \textbf{Video Games} dataset, with results shown in~\Cref{fig:hyper_combined}. The base hyperparameters are $\delta = 50$, $\gamma = -1.6$, and $\beta = 50$, with specific parameters adjusted in each plot while keeping others fixed.

$\bullet$ \textbf{Draft size ($\delta$)} controls the proportion of unseen items in recommendations, as all unseen items originate from drafting. Increasing draft size enhances inductive performance but may degrade in-sample metrics due to the fixed number of accepted candidates. The optimal value is selected via hyperparameter tuning on the validation set.

$\bullet$  \textbf{Beam size ($\beta$)} controls the search space for guided re-drafting. As shown in~\Cref{fig:hyper_combined}, increasing beam size improves in-sample performance but reduces inductive ability. 

$\bullet$ \textbf{Threshold ($\gamma$)} controls the acceptance rate of drafted candidates, impacting the number of decoding steps needed for $K$ recommendations. 
As shown in~\Cref{fig:hyper_combined}, it governs the performance-efficiency trade-off. Lower thresholds degrade in-sample performance due to overly easy candidate acceptance. We select $\gamma$ using the elbow criterion that balances marginal performance gain against additional latency.

\begin{table}[t]
\small
\centering
\setlength{\tabcolsep}{2.5mm}
\begin{tabular}{p{7mm}p{25mm}cl}
\toprule
\textbf{GR} & \textbf{Drafter} & \textbf{U-N@50} & \textbf{O-N@50} \\
\midrule

\multirow{4}{*}{TIGER}
& Baseline & -- & 0.0193 \\
& \makebox[22mm][l]{GR Encoder \scriptsize(SpecGR++)} & 0.0084 & 0.0225 {\scriptsize(+16.6\%)} \\
& Semantic-KNN & \underline{0.0085} & \underline{0.0231} {\scriptsize(+19.7\%)} \\
& UniSRec & \textbf{0.0096} & \textbf{0.0239} {\scriptsize(+23.8\%)} \\

\midrule

\multirow{4}{*}{DSI}
& Baseline & -- & 0.0198 \\
& \makebox[22mm][l]{GR Encoder \scriptsize(SpecGR++)} & \textbf{0.0061} & \underline{0.0217} {\scriptsize(+9.6\%)} \\
& Semantic-KNN & 0.0049 & \underline{0.0217} {\scriptsize(+9.6\%)} \\
& UniSRec & \underline{0.0058} & \textbf{0.0220} {\scriptsize(+11.1\%)} \\

\bottomrule
\end{tabular}
\caption{NDCG@50 on the unseen subset (U-N@50) and overall test set (O-N@50) for different GR backbones and drafter configurations on \textbf{Video Games} dataset.}
\label{tab:plug_and_play}
\end{table}

\subsection{Plug-and-Play Framework (RQ4)}
\label{sec:experiments-additional-gr-backbone}

To evaluate SpecGR’s plug-and-play capability, we integrate it with multiple inductive drafters (SemanticKNN, UniSRec, and the GR encoder (SpecGR++)) and multiple GR backbones (TIGER and DSI~\cite{tay2022transformer}). Notably, DSI employs hierarchical K-means tokenization to derive item semantic IDs. As shown in~\Cref{tab:plug_and_play}, both GR backbones are originally unable to generate unseen items. Integrating SpecGR improves overall performance by approximately 15\% on average, while enabling strong inductive recommendation. This improvement holds across lightweight retrieval (Semantic-KNN), modality-based models (UniSRec), and self-drafting using the GR encoder, demonstrating that SpecGR is robust to different drafting paradigms and input modalities. Thus, SpecGR is model-agnostic to drafter choice and serves as a plug-and-play framework that equips any semantic-ID-based GR model with inductive capability.

\section{Conclusion}
In this paper, we propose SpecGR, a plug-and-play framework that extends the capability of generative recommendation models for inductive recommendation. Our method, inspired by speculative decoding, leverages an inductive model as a drafter to propose candidate items and uses the GR model as a verifier to ensure that only high-quality candidates are recommended. We further propose two drafting strategies: (1) using an auxiliary model for flexibility, and (2) using the GR model's own encoder for parameter-efficient self-drafting. Extensive experiments on three public datasets demonstrate strong inductive and overall recommendation performance for SpecGR.

\section{Acknowledgments}
This research was partially supported by the National Sci- ence Foundation (NSF) via the grant number IIS-2432486.

\bibliography{reference}

\clearpage
\appendix

\clearpage
\appendix
\renewcommand\thesection{\Alph{section}}
\renewcommand{\thefigure}{A\arabic{figure}}
\renewcommand{\thetable}{A\arabic{table}}
\setcounter{figure}{0}   
\setcounter{table}{0}   
\setcounter{section}{0}

\section{Additional Related Work: Cold-Start Recommendation}
\label{sec:appendix-related-work}
The item cold-start problem (ICS)~\cite{zhang2025cold}, a longstanding challenge in recommender systems, requires models to recommend newly added items with limited or no historical interaction data, and therefore external knowledge have to leveraged. In what follows, we review existing works based on four major external sources:
(1) content features (2) graph relations (3) cross-domain information and (4) world knowledge from LLMs.

\paragraph{Content features}
Due to the lack of interaction records, early works naturally leveraged content features (e.g., titles, descriptions, and categories) to address cold-start recommendation. In scenarios with a few interactions, studies focused on data-efficient learning to quickly improve cold-start performance. Meta-learning~\cite{wang2020generalizing} is a widely adopted approach that pretrains models on warm interactions to capture general interaction patterns and then adapts parameters to new cold-start items using limited data. Early works achieved this through gradient-based optimization for rapid parameter adaptation~\cite{lee2019melu, finn2017model, dong2020mamo}. Later approaches incorporated task relationships to better transfer warm knowledge to cold-start scenarios using techniques such as soft-clustering~\cite{lin2021task} and domain-specific neural architectures~\cite{wu2023m2eu}.
In more challenging ICS scenarios with no interaction data, studies focus on deriving content-based item representations while still maintaining strong performance on warm items. Early works explored perturbation training strategies to reduce reliance on interactions and encourage models to learn generalized recommendations solely from content features~\cite{volkovs2017dropoutnet, zhu2020recommendation}. Others formulated item representation learning as an explicit knowledge alignment task, bridging warm interaction-based and content-based representations through methods such as contrastive learning~\cite{wei2021contrastive, zhou2023contrastive} and generative adversarial networks (GANs)~\cite{chen2022generative, alshehri2022generative}.

\paragraph{Graph relations}
In recent years, graph neural networks~\cite{wu2020comprehensive} have gained attention for modeling user-item interaction graphs in recommendation, with its power in modeling higher-order relationships beyond direct interactions and content features. However, a key challenge remains in effectively propagating information to cold-start items, where interaction signals are sparse. 
One typical methods trains the model to infer probable edges on the original use item interaction graph for cold-start items to enable information aggregation. Typical training tasks involves masking and reconstruction~\cite{kim2024content}, maximizing mutual information~\cite{wang2024mutual}, and uncertainty estimation~\cite{liu2023uncertainty}. On the other hand, another stream of work enrich the existing interaction graph with more fine-grained connections to propagate information to cold-start items. The constructed edges includes semantic links~\cite{cao2022gift, liu2020heterogeneous}, multi-view feature extractions~\cite{zheng2021multi}, and knowledge relations~\cite{du2022metakg}.

\paragraph{Cross-domain information}
To mitigate data sparsity for cold-start item recommendation, researchers have explored using cross-domain interactions to learn generalizable patterns and transfer knowledge from data-rich domains~\cite{zhu2019dtcdr, zhu2021cross, xie2022contrastive}. However, a key challenge lies in designing effective transfer techniques that account for discrepancies between domains. Early works primarily relied on overlapping information between the source and target domain, including items~\cite{singh2008relational, zhu2021cross} or attributes~\cite{tang2012cross}. To address distributional differences, distribution alignment~\cite{liu2023contrastive, wang2021low} and invariant representation learning~\cite{wu2020zero, he2018general} has been proposed to build robust item representations to facilitate knowledge transfer. Recently, with the rise of pretrained language models (PLMs), text representations have proven to be powerful semantic bridges, reducing the need for explicit entity overlaps between domains~\cite{ding2021zero, geng2022recommendation, hou2022towards}. To learn universal representations from text, many approaches leverage pretraining across multiple domains, followed by fine-tuning on domain-specific tasks to enable seamless cross-domain transfer~\cite{hou2022towards, hou2023vqrec, li2023text}.

\paragraph{World knowledge from LLMs}
In recent years, the development of Large Language Models (LLMs) \cite{zhao2023survey, achiam2023gpt, touvron2023llama} has enabled their application across a wide range of recommendation scenarios due to their well-learned world knowledge from web-scale internet data, which allow them to easily understand cold-start items. One direct application, the `LLM for Rec' methods, mainly involves prompting~\cite{liu2023chatgpt, dai2023uncovering, hou2024llmrank}, finetuning~\cite{zhang2024notellm, bao2023tallrec}, retrieval-augmented generation (RAG)~\cite{di2023retrieval, contal2024ragsys,wu2024coral}, and conversational dialogue systems~\cite{he2023large, sun2024large}. While these approaches require minimal pre-training, they suffer from high inference latency and hallucinations. To mitigate these issues, recent methods use LLMs to generate modality-based item representations\cite{hou2022towards, li2023text, hou2023vqrec} and train backbone models on these frozen representations, achieving state-of-the-art performance in cold-start scenarios.

\section{Additional Implementation Details}
\label{sec:appendix-implementation-details}

\subsection{SpecGR++ Training Details}
We train SpecGR++ with Distributed Data Parallel (DDP) using 4 GPUs (NVIDIA RTX A6000, 48GB each, Ubuntu 22.04.5 LTS). The random seed is fixed to 42 for reproducibility. During pretraining, we compute the generation loss \(\mathcal{L}_{\text{Gen}}\) with a batch size of 256, and the contrastive loss \(\mathcal{L}_{\text{CL}}\) with an effective batch size of 2048 (negatives gathered across GPUs). We train for up to 200,000 steps with early stopping, selecting learning rates from \(\{0.001, 0.0003\}\). For fine-tuning, SpecGR++ is trained for 15 epochs with a batch size of 256 and a learning rate of \(10^{-4}\). The draft threshold \(\gamma\) is tuned between -1.1 and -1.8 (step size 0.1), and beam size \(\beta\) and draft size \(\delta\) are set to 50 or \(K\). For SpecGR\textsubscript{Aux}, it is trained on a single GPU with the same hyperparameters but without distributed negative gathering.

\subsection{Embedding Batch Size Ablation}

During SpecGR++ training, to ensure the model receives sufficient negative training signals, we leverage much larger batch sizes for embedding task than generative task. We report the performance against the effective batch size, which equals the batch size $\times$ number of gpus in the distributed setting \Cref{tab:batch-size-ablation}.
We can see that larger embedding batch size positively affects the performance of embeddings from SpecGR's encoder.

\begin{table}[t]
\centering
\caption{Average validation Recall@50 for SpecGR++ across different training embedding batch sizes.}
\label{tab:batch-size-ablation}
\setlength{\tabcolsep}{0.7mm} %
\renewcommand{\arraystretch}{0.95} %
\scalebox{1.0}{ %
\begin{tabular}{lc}
\toprule
\textbf{Batch Size} & \textbf{Avg. R@50} \\ 
\midrule
1024 & 0.0598 \\ 
2048 & 0.0692 \\ 
4096 & 0.0769 \\ 
\bottomrule
\end{tabular}}
\end{table}

\subsection{Baseline Model Reproductions}
We reproduce TIGER by following the instructions provided in its original implementation~\cite{rajput2024recommender}. For Recformer, we utilize pretrained checkpoints from a popular reproduction repository\footnote{\url{https://github.com/AaronHeee/RecFormer}} and fine-tune them on our processed datasets. We implement LIGER by closely following the method details, pseudocode, and hyperparameter configurations outlined in the original paper~\cite{lin2024efficient}. For item text embeddings, we use representations obtained from a pretrained T5 sentence encoder. All other baseline models (e.g., SASRec variants and modality-based methods) are implemented using the open-source RecBole library~\cite{zhao2021recbole}. We conduct thorough hyperparameter tuning for all baseline models and report results using the best-performing configuration for fair comparison.

\subsection{TIGER\textsubscript{C} Implementation}
\label{sec:appendix-tiger-cold-start}

Following~\citet{rajput2024recommender}, during inference, we use TIGER to generate top-$K$ candidates and retrieve unseen items whose semantic prefix (\ie semantic IDs without the last identifier token) appears in this list. A hyperparameter \(\epsilon\) controls the maximum proportion of unseen items included. After retrieving $x$ unseen candidates, where $x <= \epsilon \cdot K$, we append them after the first $K - x$ generated candidates to finalize the recommendation.  We tune \(\epsilon\) on each dataset and report the best results in \Cref{tab:ablation-main}.

\subsection{SpecGR++ Training Details}

\begin{table}[t]
\centering
\caption{Statistics of the datasets. ``New\%'' denotes the proportion of interactions with unseen target items. ``\#Items'' and ``\#Inter.'' are in thousands (K).}
\label{table:datasets}
\setlength{\tabcolsep}{2.2pt}
\small
\begin{tabular}{l
>{\raggedleft\arraybackslash}p{0.9cm}
>{\raggedleft\arraybackslash}p{0.8cm}
>{\raggedleft\arraybackslash}p{0.9cm}
>{\raggedleft\arraybackslash}p{0.9cm}
>{\raggedleft\arraybackslash}p{0.8cm}
>{\raggedleft\arraybackslash}p{0.9cm}
>{\centering\arraybackslash}p{0.8cm}}
\toprule
\multirow{2}{*}{\textbf{Dataset}} 
  & \multicolumn{2}{c}{\textbf{Items}} 
  & \multicolumn{1}{c}{\textbf{Train}} 
  & \multicolumn{2}{c}{\textbf{Valid}} 
  & \multicolumn{2}{c}{\textbf{Test}} \\
\cmidrule(lr){2-3}
\cmidrule(lr){4-4}
\cmidrule(lr){5-6}
\cmidrule(lr){7-8}
& \#Items & New\%
& \#Inter.
& \#Inter. & New\%
& \#Inter. & New\% \\
\midrule
\textbf{Games}  
  & 25.6 & 10.3
  & 645.3 & 33.1 & 27.9
  & 41.5 & 60.3 \\
\textbf{Office} 
  & 77.6 & 15.1
  & 1230.2 & 136.1 & 16.2
  & 211.3 & 59.4 \\
\textbf{Phones} 
  & 111.5 & 15.1
  & 1841.5 & 232.9 & 33.0
  & 297.4 & 68.3 \\
\bottomrule
\end{tabular}
\end{table}

\section{Case Study: Guided Redrafting}
\label{appendix-guided-redrafting}

From \Cref{tab:ablation-main}, guided re-drafting leverages the generative backbone to refine redrafted candidates, enhancing overall performance. To analyze its mechanism and impact, we conduct a case study under a fixed threshold ($\gamma = -1.4$). First, \Cref{tab:acceptance_rates} reports the comparison of acceptance rate per decoding step between naive drafting (w/o) and guided re-drafting (w). Under a strict threshold, we can see that the step-1 acceptance rate is low, highlighting the need for re-drafting. Guided re-drafting consistently increases the number of accepted items in later steps, improving verification efficiency. Furthermore, \Cref{tab:performance_redrafting} presents the impact of guided re-drafting on overall and subset-specific performance under a fixed threshold. As shown, guided re-drafting yields improvements across all categories, particularly in the unseen subset, where recall and NDCG scores rise by approximately 10\%. This validates that leveraging redrafted tokens to constrain the drafting scope results in higher-quality candidates in subsequent steps.

\begin{table}[t]
\centering
\setlength{\abovecaptionskip}{2pt} %
\setlength{\belowcaptionskip}{2pt} %
\caption{Acceptance rate (\%) per decoding step with and without guided re-drafting when $\gamma = -1.4$.}
\label{tab:acceptance_rates}
\setlength{\tabcolsep}{1.2mm} %
\scalebox{0.92}{ %
\begin{tabular}{lcccc}
\toprule
\textbf{Step} & \textbf{1} & \textbf{2} & \textbf{3} & \textbf{4} \\ 
\midrule
w/o & 10.34 & 5.93 & 4.25 & 2.63 \\  
w & 10.34 & 6.62 & 5.51 & 5.75 \\  
\bottomrule
\end{tabular}}
\end{table}

\begin{table}[t]
\centering
\setlength{\abovecaptionskip}{2pt} %
\setlength{\belowcaptionskip}{2pt} %
\caption{Performance comparison with and without guided re-drafting when $\gamma = -1.4$.}
\label{tab:performance_redrafting}
\setlength{\tabcolsep}{1.2mm} %
\scalebox{0.92}{ %
\begin{tabular}{lcccccc}
\toprule
\multirow{2}{*}{} & \multicolumn{2}{c}{\textbf{Overall}} & \multicolumn{2}{c}{\textbf{Unseen}} & \multicolumn{2}{c}{\textbf{In-Sample}} \\ 
\cmidrule(lr){2-3} \cmidrule(lr){4-5} \cmidrule(lr){6-7}  
& \textbf{R@50} & \textbf{N@50} & \textbf{R@50} & \textbf{N@50} & \textbf{R@50} & \textbf{N@50} \\ 
\midrule
w/o & 0.0702 & 0.0218 & 0.0288 & 0.0076 & 0.1331 & 0.0434 \\  
w & \textbf{0.0721} & \textbf{0.0223} & \textbf{0.0317} & \textbf{0.0082} & \textbf{0.1333} & \textbf{0.0436} \\  
\bottomrule
\end{tabular}}
\end{table}

\begin{table}[t]
\small
\centering
\caption{Performance comparison of SpecGR\textsubscript{Aux} (UniSRec drafter) against ensemble variants. The best and second-best scores are bolded and underlined.}
\label{tab:ablation-ensemble}
\resizebox{\columnwidth}{!}{%
\setlength{\tabcolsep}{1.0mm}
\begin{tabular}{lcccccc}
\toprule
\multirow{2}{*}{\textbf{Model}} &
\multicolumn{2}{c}{\textbf{Overall}} &
\multicolumn{2}{c}{\textbf{Unseen}} &
\multicolumn{2}{c}{\textbf{In-Sample}} \\ 
\cmidrule(l){2-3}\cmidrule(l){4-5}\cmidrule(l){6-7}
& \textbf{R@50} & \textbf{N@50} & \textbf{R@50} & \textbf{N@50} & \textbf{R@50} & \textbf{N@50} \\ 
\midrule
\multicolumn{7}{c}{\emph{Single Model}}\\
\midrule
TIGER                    & 0.0584 & 0.0193 &      -- &      -- & 0.1472 & \underline{0.0486} \\  
UniSRec                  & 0.0621 & 0.0200 & \underline{0.0118} & \underline{0.0029} & 0.1386 & 0.0461 \\  
\midrule
\multicolumn{7}{c}{\emph{Ensemble Method}}\\
\midrule
Score-based              & 0.0571 & 0.0191 & 0.0050 & 0.0009 & 0.1333 & 0.0456 \\  
Ranking-based            & \underline{0.0678} & \underline{0.0218} & 0.0056 & 0.0011 & \textbf{0.1624} & \textbf{0.0532} \\  
2-Stage                  & 0.0621 & 0.0205 & \underline{0.0118} & 0.0026 & 0.1386 & 0.0477 \\  
\midrule
SpecGR (UniSRec)         & \textbf{0.0778} & \textbf{0.0239} & \textbf{0.0312} & \textbf{0.0096} & \underline{0.1485} & 0.0457 \\  
\bottomrule
\end{tabular}}
\end{table}

\section{Ablation: Comparison with Ensemble Methods}
\label{sec:appendix-ablation-ensemble}
We compare SpecGR\textsubscript{Aux} with ensemble-based variants combining GR and modality-based models, as shown in~\Cref{tab:ablation-ensemble}. The \textit{Score-based Ensemble} linearly combines TIGER's likelihood scores and UniSRec's ranking scores. The \textit{Ranking-based Ensemble} averages item ranking positions from both models to mitigate score scale mismatches. The \textit{2-Stage Ensemble} selects top-$K$ items using UniSRec and re-ranks them with TIGER's scores. All ensemble methods have unseen performance bounded by the two base models, TIGER and UniSRec, since simple score aggregation or re-ranking does not introduce inductive capabilities. In contrast, SpecGR uses guided re-drafting to effectively leverage the backbone’s modeling capability for enhancing unseen candidates' quality. The results illustrate the effectiveness of the proposed SpecGR methods.

\section{Inference Speed Acceleration}
\label{sec:appendix-inference-acc}

Generative recommendation (GR) models rely on autoregressive next-token generation, which incurs high inference latency. SpecGR addresses this by employing speculative retrieval-based drafting, where candidate items proposed by the drafter can be efficiently verified by the GR verifier, significantly reducing the number of required autoregressive decoding steps.

\subsection{Drafting and Verification Latency}
\label{sec:appendix-latency-breakdown}
To clearly illustrate SpecGR's efficiency, we report detailed inference latency results (in seconds) separated into drafting (D) and verification (V) stages across various draft sizes. As shown in \Cref{tab:latency-detailed}, SpecGR++ achieves remarkably low drafting latency due to its efficient encoder-based drafting mechanism, demonstrating significant overall speed improvements compared to TIGER.

\begin{table}[h]
    \centering
    \caption{Empirical inference latency (seconds) across drafting (D) and verification (V) stages.}
    \label{tab:latency-detailed}
    \vspace{0.5em}
    \resizebox{\columnwidth}{!}{%
    \setlength{\tabcolsep}{1.5mm}
    \begin{tabular}{lccc}
        \toprule
        \textbf{Draft Size} & \textbf{20 (D/V)} & \textbf{50 (D/V)} & \textbf{100 (D/V)} \\
        \midrule
        SpecGR (UniSRec) & 0.0110 / 0.0261 & 0.0112 / 0.0249 & 0.0112 / 0.0227 \\
        SpecGR++ & \textbf{0.0003} / 0.0261 & \textbf{0.0003} / 0.0249 & \textbf{0.0003} / 0.0235 \\
        TIGER & - / 0.0403 & - / 0.0403 & - / 0.0403 \\
        \bottomrule
    \end{tabular}
    }
\end{table}

\begin{figure}[t]
  \centering
  \includegraphics[width=1.0\columnwidth]{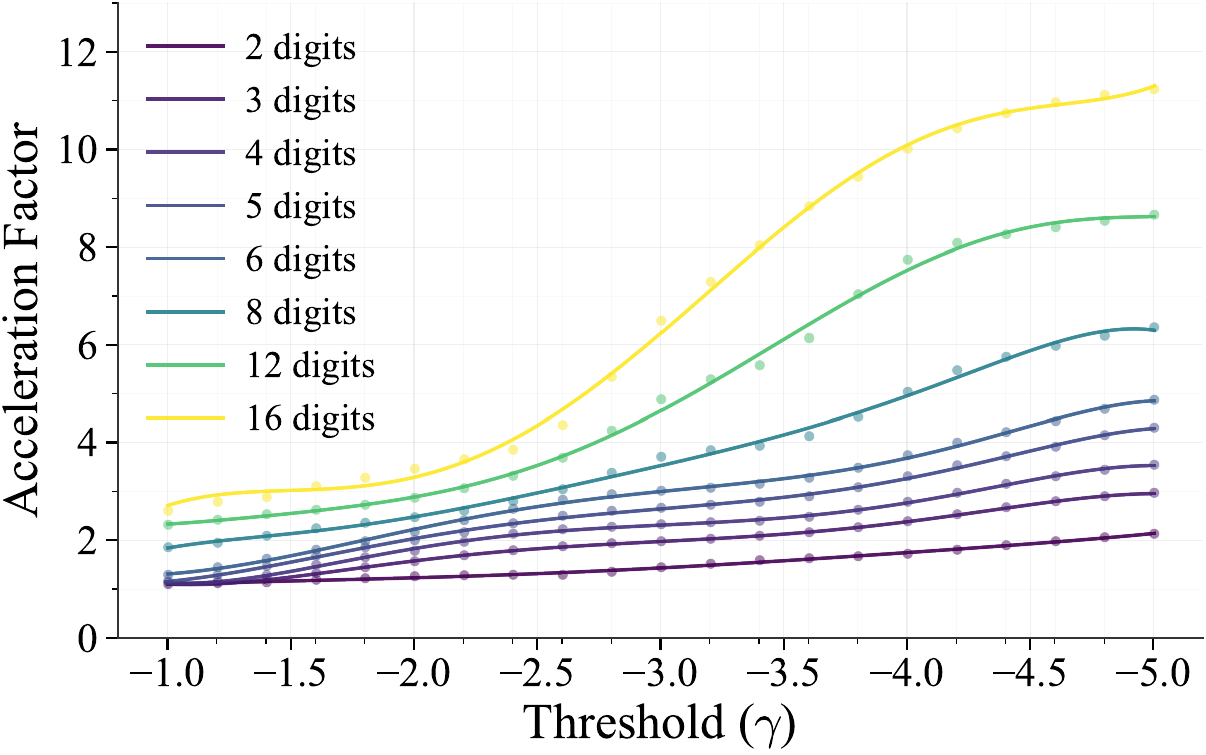}
  \caption{Inference speed acceleration factor \wrt different numbers of semantic ID digits.}
  \label{fig:speed-vs-codebook-length}
\end{figure}

\subsection{Acceleration w.r.t. Generation Length}
\label{sec:appendix-latency-generation-length}
We analyze how SpecGR reduces inference latency as the generation length increases. Specifically, we illustrate the acceleration factor \wrt the number of digits in semantic IDs. We train SpecGR++ with a TIGER backbone across different semantic ID lengths (ranging from 2 to 16 digits) and report the acceleration factor relative to the TIGER baseline. The acceleration factor is defined as the inverse ratio of inference time. As shown in~\Cref{fig:speed-vs-codebook-length}, in standard GR settings with 4-digit semantic IDs, SpecGR achieves a $1.7\times$ inference speedup while maintaining strong overall performance (see the Experiments section). Moreover, longer semantic IDs lead to even higher acceleration as fewer decoding steps are needed per item.

\subsection{Theoretical Time Complexity}
\label{sec:appendix-latency-complexity}
We further analyze the expected time complexity of SpecGR, where each item is represented by $l$ semantic tokens. Let $K$ be the number of items to recommend, $\delta$ the draft size per iteration, and $p$ the acceptance probability of a drafted item. We denote the runtime for a single forward pass in the GR encoder and decoder as $C_{encoder}$ and $C_{decoder}$, respectively. A standard GR model (e.g., TIGER) must decode all $l$ tokens for $K$ beams, yielding a complexity:
\[
\text{Cost}_{\text{TIGER}} = \mathcal{O}(C_{encoder} + l \cdot C_{decoder}).
\]
For SpecGR, inductive drafting requires one forward pass and a KNN search, denoted as $C_{draft}$. Verification and beam search happen in parallel with a cost of $C_{decoder}$ per iteration. The expected number of iterations to obtain $K$ valid recommendations is bounded by:
\[
T = \mathcal{O}\left(\frac{K}{\delta p}\right) \leq \mathcal{O}(l).
\]
Thus, SpecGR's expected time complexity is:
\[
\text{Cost}_{\text{SpecGR}} = \mathcal{O}(C_{draft} + C_{encoder} + T \cdot C_{decoder}),
\]
with an expected acceleration factor:
\[
\frac{C_{encoder} + l \cdot C_{decoder}}{C_{draft} + C_{encoder} + \frac{K}{\delta p} \cdot C_{decoder}}.
\]

\subsection{Empirical Acceptance Rates and Speedup}
\label{sec:appendix-latency-empirical}
We empirically measure the acceptance probability $p$ for various drafting methods and report their corresponding acceleration factors in \Cref{tab:acceleration}. For the SemanticKNN drafter, item sequences are encoded using mean text embeddings from the most recent five items, and top-$K$ items are retrieved via KNN search. Across all methods, SpecGR consistently outperforms traditional GR models.

\begin{table}[h]
    \centering
    \caption{Empirically measured acceptance rate $p$ and expected acceleration factor for different drafters at $\gamma = -1.4$.}
    \label{tab:acceleration}
    \vspace{0.5em}
    \small
    \resizebox{\columnwidth}{!}{%
    \setlength{\tabcolsep}{1.5mm}
    \begin{tabular}{lcc}
        \toprule
        \textbf{Drafter} & \textbf{Acceptance Rate ($p$)} & \textbf{Acceleration Factor} \\
        \midrule
        SpecGR++ Encoder & 0.44 & 1.72 \\
        UniSRec & 0.35 & 1.38 \\
        SemanticKNN & 0.20 & 1.11 \\
        \bottomrule
    \end{tabular}
    }
\end{table}

\subsection{Scalability Analysis}
\label{sec:appendix-latency-scalability}
SpecGR exhibits favorable scalability in multiple dimensions compared to traditional GR models:
\begin{itemize}
    \item \textbf{Model Size.} As the GR model scales, assuming $C_{encoder} \approx C_{decoder}$, SpecGR maintains a constant acceleration factor $\frac{l+1}{\frac{K}{\delta p}+1}$, demonstrating robust scalability.
    \item \textbf{Semantic ID Length.} The acceleration factor grows linearly as the semantic ID length increases, i.e., $\mathcal{O}\left(\frac{l \cdot \delta \cdot p}{K}\right)$, highlighting SpecGR's advantage in scenarios requiring long sequences, as quantitatively shown in \Cref{fig:speed-vs-codebook-length}.
\end{itemize}

\section{Additional Further Analysis}
\label{sec:appendix-additional-further-analysis}

We briefly discuss SpecGR’s capabilities compared to traditional methods. Leveraging the draft-then-verify framework, SpecGR seamlessly integrates into diverse recommendation scenarios, adapting to dynamic needs (\Cref{tab:modality-comparison}). For instance, when deployed as a ranker model, it exploits autoregressive generation for efficient subset ranking. Additionally, its tunable hyperparameters enable real-time controllable inductive ability and controllable inference speed. See the section Other New Capabilities for broader implications.

\subsection{Subset Ranking}
\label{sec:appendix-subset-ranking}
In this section, we assess new capability of SpecGR in efficient subset ranking compared to traditional subset ranking methods for GR models. 
Since most deep learning-based recommendation methods unavoidably introduce high latency, they are typically used as ranking models rather than retrieval models~\cite{covington2016deep,hou2024llmrank}.
In the subset ranking setting, the model of interest is applied as a ranker to rank a given subset of items $(\mathcal{I}_r \subset \mathcal{I}, |\mathcal{I}_r| \ll |\mathcal{I}|)$.
Subset ranking is out-of-the-box for traditional sequential recommendation models by selecting a subset of candiates when performing KNN search. However, for generative recommendation models, subset ranking presents a challenge as they inherently recommend items by searching for the top \( K \) decoding paths across the entire item space. 

\begin{figure}[t]
    \centering
    \begin{minipage}[t]{0.48\linewidth}
        \centering
        \includegraphics[width=\linewidth]{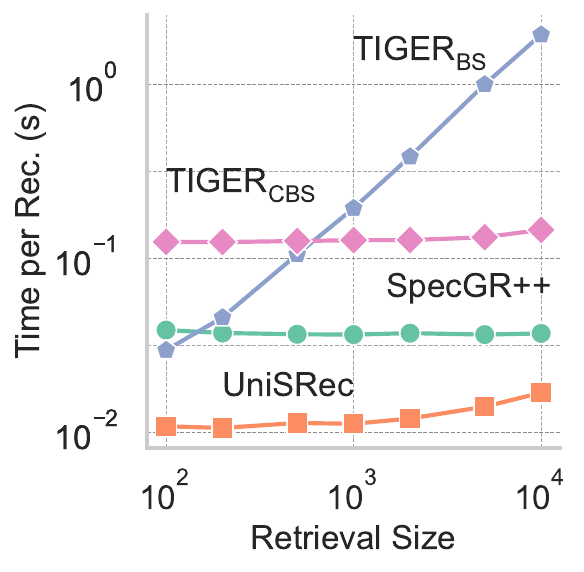}
        \label{fig:subset-ranking}
    \end{minipage}
    \hfill
    \begin{minipage}[t]{0.48\linewidth}
        \centering
        \includegraphics[width=\linewidth]{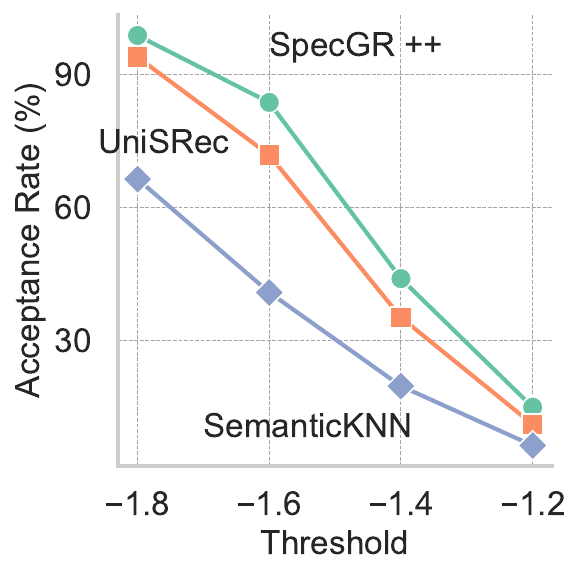}
        \label{fig:acceptance-rate}
    \end{minipage}
    \caption{(Left) Inference latency comparison for subset ranking. Both x- and y-axis use log scale. (Right) Acceptance rate comparison for different drafting strategies.}
    \label{fig:combined-figure}
\end{figure}

\paragraph{Batch scoring (BS)} A simple approach to address this is through batch scoring, which involves splitting the item subset into fixed-size batches and scoring them consecutively with the generative model. However, this method grows linearly with batch size and is impractical for large subsets, as shown in~\Cref{fig:combined-figure}. 

\paragraph{Constrained beam search (CBS)} An enhanced method is to use constrained beam search \cite{anderson2016guided, post2018fastlexicallyconstraineddecoding, de2020autoregressive}. Constrained beam search constructs a trie using all allowed prefixes to restrict the search space at each decoding step. However, this approach introduces significant computational overhead. The time complexity for CBS is $\mathcal{O}(T \cdot n_b \cdot 2^C)$, where $T$ is the sequence length, $n_b$ is the beam size, and $C$ is the number of constraints~\cite{chen2025reinforcement}. The large computational overhead and exponential growth with the number of constraints makes it inefficient for large-scale ranking tasks, as shown quantitatively in~\Cref{fig:combined-figure}.

\paragraph{SpecGR for subset ranking} SpecGR effectively addresses this issue by restricting the drafter model's range to a specified subset. This ensures that all recommendations originate from within the subset. As demonstrated in~\Cref{fig:combined-figure}, SpecGR achieves a \( 3.5 \times \) speedup for subset sizes \( < 10^4 \) compared to TIGER with constrained beam search (denoted as CBS). Moreover, SpecGR maintains a time complexity that is bounded by its full ranking complexity as the retrieval size increases, making it a highly efficient solution for subset ranking tasks.

\begin{table}[t]
\small
\centering
\caption{Comparison of different models based on parameter efficiency and training time.}
\label{tab:model_params}
\scalebox{0.92}{
\setlength{\tabcolsep}{0pt} %
\begin{tabular}{@{}l>{\centering\arraybackslash}p{1.5cm}>{\centering\arraybackslash}p{1.5cm}>{\centering\arraybackslash}p{1.5cm}>{\centering\arraybackslash}p{1.5cm}>{\centering\arraybackslash}p{1.5cm}@{}}
\toprule
\multirow{2}{*}{\textbf{Model}} & \multicolumn{3}{c}{\textbf{Trainable (M)}} & \multirow{2}{*}{\makecell{\textbf{Non-train} \\ \textbf{-able (M)}}} & \multirow{2}{*}{\makecell{\textbf{Training} \\ \textbf{Time (h)}}} \\ \cmidrule(lr){2-4}
 & \makecell{\textbf{Total}} & \makecell{\textbf{Non-emb}} & \makecell{\textbf{Emb}} & & \\ \midrule
SASRec\textsubscript{ID} & 7.24 & 0.10 & 7.13 & 0 & 3.6 \\
UniSRec & 2.90 & 2.90 & 0 & 85.62 & 18.3 \\
Recformer & 233.73 & 106.32 & 127.41 & 0 & 226.0 \\
TIGER & 13.26 & 13.11 & 0.15 & 0 & 16.2 \\
TIGER\textsubscript{C} & 13.26 & 13.11 & 0.15 & 0 & 16.2 \\
SpecGR\textsubscript{Aux} & 16.16 & 16.02 & 0.15 & 85.62 & 34.5 \\
SpecGR++ & 13.28 & 13.13 & 0.15 & 14.27 & 42.8 \\ \bottomrule
\end{tabular}}
\end{table}

\begin{table}[t]
\centering
\caption{Comparison of SpecGR against existing models across different scenarios and model capabilities.}
\label{tab:modality-comparison}
\scriptsize
\setlength{\tabcolsep}{0mm}
\begin{tabular}{@{}l
>{\centering\arraybackslash}p{1.4cm}
>{\centering\arraybackslash}p{1.4cm}
>{\centering\arraybackslash}p{1.4cm}
>{\centering\arraybackslash}p{1.4cm}
>{\centering\arraybackslash}p{1.4cm}@{}}
\toprule
\multirow{2}{*}{\textbf{Model}} &
\multicolumn{2}{c}{\textbf{Recommendation Scenario}} &
\multicolumn{3}{c}{\textbf{Model Capability}} \\
& \makecell[c]{Efficient \\ Subset \\ Ranking} 
& \makecell[c]{Inductive \\ Recom- \\ mendation} 
& \makecell[c]{Auto- \\ regressive \\ Generation} 
& \makecell[c]{Controllable \\ Inductive \\ Ability} 
& \makecell[c]{Controllable \\ Inference \\ Speed} \\
\midrule
SASRec\textsubscript{ID}     
& \textcolor{teal}{\CheckmarkBold}
& \textcolor{purple}{\XSolidBrush}
& \textcolor{purple}{\XSolidBrush}
& \textcolor{purple}{\XSolidBrush}
& \textcolor{purple}{\XSolidBrush} \\
UniSRec                     
& \textcolor{teal}{\CheckmarkBold}
& \textcolor{teal}{\CheckmarkBold}
& \textcolor{purple}{\XSolidBrush}
& \textcolor{purple}{\XSolidBrush}
& \textcolor{purple}{\XSolidBrush} \\
Recformer                   
& \textcolor{teal}{\CheckmarkBold}
& \textcolor{teal}{\CheckmarkBold}
& \textcolor{purple}{\XSolidBrush}
& \textcolor{purple}{\XSolidBrush}
& \textcolor{purple}{\XSolidBrush} \\
TIGER                       
& \textcolor{purple}{\XSolidBrush}
& \textcolor{purple}{\XSolidBrush}
& \textcolor{teal}{\CheckmarkBold}
& \textcolor{purple}{\XSolidBrush}
& \textcolor{purple}{\XSolidBrush} \\
TIGER\textsubscript{C}      
& \textcolor{purple}{\XSolidBrush}
& \textcolor{teal}{\CheckmarkBold}
& \textcolor{teal}{\CheckmarkBold}
& \textcolor{teal}{\CheckmarkBold}
& \textcolor{purple}{\XSolidBrush} \\
\textbf{SpecGR\textsubscript{Aux}} 
& \textcolor{teal}{\CheckmarkBold}
& \textcolor{teal}{\CheckmarkBold}
& \textcolor{teal}{\CheckmarkBold}
& \textcolor{teal}{\CheckmarkBold}
& \textcolor{teal}{\CheckmarkBold} \\
\textbf{SpecGR++}           
& \textcolor{teal}{\CheckmarkBold}
& \textcolor{teal}{\CheckmarkBold}
& \textcolor{teal}{\CheckmarkBold}
& \textcolor{teal}{\CheckmarkBold}
& \textcolor{teal}{\CheckmarkBold} \\
\bottomrule
\end{tabular}
\end{table}

\subsection{Other New Capabilities}
\label{sec:appendix-new-capabilities}

As a direct impact of the adjustable hyperparameters analyzed in the Experiments section, SpecGR possesses two other new capabilities: controllable inductive ability and controllable inference speed.

\paratitle{Controllable inductive ability} allows platforms to dynamically adjust their recommendation strategy based on seasonal demand, favoring new items during certain periods while prioritizing established products at others. For instance, e-commerce platforms could increase the inductive ability of SpecGR
to promote new items, and reduce it during clearance sales.

\paratitle{Controllable inference speed} enables platforms to trade off model performance for faster inference speed during high-traffic periods, ensuring responsive user experiences.

In summary, as shown in \Cref{tab:modality-comparison}, SpecGR inherits the architecture of generative recommendation, allowing effective scaling on large datasets. It also extends high performance and low inference speed to broader real-life recommendation settings, adapting to specific data characteristics and recommendation needs.

\section{Further discussion on SpecGR++ Architecture}
\label{sec:appendix-discussion-on-specGR-design}
SpecGR++ utilizes its encoder as the self-drafter, effectively eliminating the need for maintaining a separate draft model for drafting, resulting in a more integrated method during inference. In this section, we will study the additional advantages of the SpecGR++ compared to the SpecGR with an auxiliary model.

\subsection{Parameter Efficiency and Speed}
\label{sec:appendix-parameter-efficiency-and-speed}
First, SpecGR++ uses intermediate encoder outputs for drafting, resulting in nearly no additional computational cost and parameter sizes compared to GR. We report the total number of parameters and training time required for different methods in \Cref{tab:model_params}. As we can see, SpecGR inherents GR's advantages for scaling for the larger dataset as it assigns most of the parameters into non-embedding layers. Due to the additional embedding training tasks, SpecGR++ training time is slightly longer than training a drafter model and TIGER, and is 2.6x more GPU hours than training a TIGER. However, we believe that it is worthwhile to consider the acceleration during inference time. We've also provided a distributed training implementation in the released code. 

\subsection{Unified Representation Space} 
\label{sec:appendix-unified-representation-space}
Because both the drafter and verifier use the encoder's representation, we observe a higher acceptance rate for the self-drafter compared to drafting with auxiliary models. This leads to better recommendation efficiency, where less decoding step is required.

Notably, we also observe a slight increase in generative performance after SpecGR++ pretraining compared to generation-only training (\eg TIGER). Recent studies have shown similar results, indicating that with a unified representation space, a model can maintain high performance in both generative and embedding tasks in NLP \cite{muennighoff2024generative}. Our study further confirms that generation and representation are not conflicting tasks in the recommendation setting but rather two complementary approaches to solving the same problem. We look forward to seeing future research in recommendation systems that explores the unification and overlap between generative recommendation and representational recommendation.

\end{document}